# Magnetodipolar interlayer interaction effect on the magnetization dynamics of a trilayer square element with the Landau domain structure

D.V. Berkov, N.L. Gorn

*Innovent Technology Development, Prüssingstr. 27b, D-07745 Jena, Germany*

## ABSTRACT

We present a detailed numerical simulation study of the effects caused by the magnetodipolar interaction between ferromagnetic (FM) layers of a trilayer magnetic nanoelement on its magnetization dynamics. As an example we use a Co/Cu/Ni$_{80}$Fe$_{20}$ element with a square lateral shape where the magnetization of FM layers forms a closed Landau-like domain pattern. First we show that when the thickness of the non-magnetic (NM) spacer is in the technology relevant region $h \sim 10$ nm, magnetodipolar interaction between 90$^\text{o}$ Neel domain walls in FM layers *qualitatively* changes the *equilibrium* magnetization state of these layers. In the main of the paper we compare the magnetization *dynamics* induced by a sub-nsec field pulse in a single-layer Ni$_{80}$Fe$_{20}$ (Py) element and in the Co/Cu/Py *trilayer* element. Here we show that (*i*) due to the spontaneous symmetry breaking of the Landau state in the FM/NM/FM trilayer its domains and domain walls oscillate with different frequencies and have different spatial oscillation patterns; (*ii*) magnetization oscillations of the trilayer domains are strongly suppressed due to different oscillation frequencies of domains in Co and Py; (*iii*) magnetization dynamics qualitatively depends on the relative rotation sense of magnetization states in Co and Py layers and on the magnetocrystalline anisotropy kind of Co crystallites. Finally we discuss the relation of our findings with experimental observations of magnetization dynamics in magnetic trilayers, performed using the element-specific time-resolved X-ray microscopy.

# I. INTRODUCTION

Studies of nanoelements patterned out of magnetic multilayers constitute now a rapidly growing research area due to their already existing and very promising future applications. Using modern patterning technologies it is possible to produce arrays of nanoelements with lateral sizes of ~ $10^2$ - $10^3$ nm and nearly arbitrary shapes. Such elements and their arrays can be employed in magnetic random access memory (MRAM) cells, miniaturized magnetoresistance sensors (in read/write hard disk heads), advanced high-density storage media, spintronic devices [1] etc.

Small lateral sizes of these single- and multilayered structures lead to qualitatively new features of their magnetization dynamics, with the quantization of their spin wave eigenmodes being the most famous example (see, e.g., [2, 3, 4, 5]). Thorough understanding of this novel features is crucially important both for the progress of the fundamental research in this area and for the development of reliable high-technology products based on such systems.

In the last decade extensive experimental and theoretical effort was dedicated to the studies of magnetization dynamics of *single*-layer nanoelements. Among them the nanodisks possessing closed magnetization configuration with the central vortex (for continuous disks) or without it (for rings with the hole in the middle) represent the simplest non-trivial example due to their circular form and hence - axially symmetric magnetization configuration. Magnetization dynamics of these nanodisks has been extensively studied using advanced experimental techniques, analytical theories and numerical simulations [6] and is satisfactory understood.

The next complicated case is a square or rectangular single-layer nanoelement with either saturated magnetization state or closed Landau domain structure. In the state close to saturation the main non-trivial effect is due to the strong demagnetizing field near the element edges perpendicular to the field and magnetization direction; corresponding dynamics could also be understood quantitatively combining experimental and theoretical methods (see, e.g., [2, 4] and references therein). The closed Landau magnetization pattern is much more demanding at least from the theoretical point of view, because magnetization dynamics of this structure exhibits both highly localized (oscillation of the central core and domain walls) and extended (oscillation of domain areas) modes. But many important features of this dynamics could be also understood very recently using such advanced experimental methods as time-resolved Kerr microscopy and space resolved quasielastic Brilloin light scattering techniques, supported by detailed numerical simulations [7, 8, 9, 10, 11].

However, single-layer elements are not very interesting from the point of view of potential applications, because almost any technical device based on magnetic nanolayers employs - for various, but fundamental reasons - mainly multilayer structures. An additional layer (or several such layers) is required, e.g., as a reference layer with 'fixed' magnetization to detect via some MR effect the resistance change when the 'free' layer changes its magnetization direction, or as an electron spin polarizer in spintronic devices, etc. For this reason the magnetization dynamics of *multilayered* structures is of the major interest.

In such structures the interlayer interaction effects play often a very important role. Even if we leave aside a strong exchange (RKKY) coupling present in structures with very thin non-magnetic spacers consisting of some specific materials like Ru, we are still left with the unavoidable magnetodipolar coupling between the layers. This coupling is identically zero only for a multilayer structure with infinitely extended and homogeneously magnetized layers - the situation which virtually never is encountered in practice. For this reason understanding of the magnetodipolar interlayer interaction influence is absolutely necessary for further progress.

This interaction is especially strong in situations, where the layer magnetization is - at least in some regions - perpendicular to the free layer surface, thus inducing very large "surface magnetic charges" and consequently - high stray fields. Typical example of such systems are multilayers with a perpendicular magnetic anisotropy and nanoelements where the magnetization lies in the layer plane, but is nearly saturated, so that large stray fields emerge near the side edges of a multilayer stack. The influence of the interlayer interaction in such systems has been extensively investigated in the past for the quasistatic magnetization structures (see, e.g., the review [12] and references therein) and very recently - by studies of the magnetization dynamics [13].

To avoid the strong interaction caused by the magnetization directed normally to the free surface, a commonly used idea is to employ multilayer nanoelements with closed magnetization structures. In nanodisks such a structure is represented by an in-plane rotating magnetization, containing a central vortex as the only element producing strong demagnetizing field. Magnetodipolar interaction between the cores of such vortices in such multilayered circular nanodots has been investigated recently in [13].

For the next commonly used nanoelement shape - magnetic rectangle - the closed magnetization structure is achieved by the famous Landau pattern with four homogeneously magnetized domains and four $90^o$ domain walls (in case of a square element). For sufficiently thin films commonly used in technologically relevant systems these walls are the so called Neel walls [14] with the magnetization lying almost in the element plane. Such a magnetization configuration contains only volume 'magnetic charges' (no free poles on the element surface, and hence - no surface charges), which are usually weaker than surface charges. For this reason the interlayer interaction mediated by the domain walls is multilayer elements with the closed magnetization structure is expected to be weaker than in multilayer stacks with saturated magnetization.

However, recently several research groups [15, 16, 17, 18] have demonstrated that the stray field caused by the magnetization of vortex and/or Neel walls in one layer of a multilayered system can still strongly affect the magnetic state of other layers. Most of these studies use simplified DW models which enable a semianalytical treatment of the problem (see, e.g., [15, 16]), but rigorous micromagnetic simulations have confirmed that the stray field of a vortex wall with the Neel cap [17] or of a purely Neel wall [18] can strongly affect the magnetic state of other layers even if the interlayer separation is as large as ~ 100 nm [18].

In this study we consider multilayer square elements with lateral sizes ~ 1 mkm and thicknesses of magnetic layers and non-magnetic spacers ~ 10 nm (i.e., geometry typical for numerous applications). We shall demonstrate that in such systems the magnetodipolar interlayer interaction due to the Neel domain walls of the closed Landau structure is strong enough to change qualitatively both the *quasistatic* magnetization structure and magnetization *dynamics* of a system. The paper is organized as follows. After the brief description of our simulation methodology (Sec. *II.A*) we show reference results for a square single-layer element which will serve for comparison with multilayered systems. Afterwards we analyze in detail the effect of the interlayer interaction on the quasistatic magnetization structure and magnetization dynamics for a square FM/NM/FM trilayer (Sec. *II.C*), considering both the influence of the initial magnetization state - compare Sec. *II.C* and *D*, and the effect of the random polycrystalline grain structure - compare Sec. *II.D* and *E*. In Sec. *III* we compare our results with (unfortunately very few) available experimental and numerical studies of similar systems, and discuss the possibility of an experimental verification of our simulation predictions.

## II. NUMERICAL SIMULATION RESULTS

### A. Numerical simulations setup

In this study we have simulated the trilayer element Co/Cu/Py with the following geometry shown in Fig. 1 (when not stated otherwise): lateral sizes 1 x 1 mkm$^2$, Co and Py layer thicknesses $h_{Co} = h_{Py} = 25$ nm, Cu interlayer thickness (non-magnet spacer thickness between Co and Py layers) $h_{sp} = 10$ nm. Both magnetic layers were discretized into $N_x$ x $N_z$ x $N_y$ = 200 x 200 x 4 rectangular prismatic cells. We have checked that the discretization into at least 4 in-plane sublayers was necessary to reproduce correctly the 3D magnetization structure of 90$^o$ domain walls (DW) present in the equilibrium Landau magnetization state of square nanoelements.

The following magnetic parameters have been used: for Py - saturation magnetization $M_S^{Py} = 860$ G, exchange stiffness constant $A^{Py} = 1 \cdot 10^{-6}$ erg/cm and cubic magnetocrystalline grain anisotropy $K_{cub}^{Py} = 5 \cdot 10^3$ erg/cm$^3$; for Co - $M_S^{Co} = 1400$ G, $A^{Co} = 3 \cdot 10^{-6}$ erg/cm and magnetocrystalline grain anisotropies $K_{cub}^{Co} = 6 \cdot 10^5$ erg/cm$^3$ for the cubic *fcc* modification of Co and $K_{un}^{Co} = 4 \cdot 10^6$ erg/cm$^3$ for its uniaxial *hcp* modification (see [19] for discussion of all values for Co). The average grain (crystallite) size $\langle D \rangle = 10$ nm (in all directions) with 3D randomly oriented anisotropy axes of various crystallites was used for both magnetic materials. There was no correlation between the crystallites in Py and Co layers.

Simulations of both the equilibrium magnetization structure and magnetization dynamics were performed using our commercially available MicroMagus package (see [20] for implementation details). For simulations of the magnetization dynamics the package employs the optimized Bulirsch-Stoer method with the adaptive step-size control to integrate the Landau-Lifshitz-Gilbert equation for the magnetization motion with the standard linear Gilbert damping (damping constant was set to $\lambda = 0.01$ throughout our simulations). Due to the small amplitude of magnetization oscillations studied here we believe that this simplest damping form adequately describes the energy dissipation in our system. We also did not take into studies additional damping caused by the spin pumping effect in magnetic multilayers (see, e.g., [21]), because this is beyond the scope of this paper.

We have studied magnetization dynamics of our system in a *pulsed* magnetic field applied *perpendicularly* to the element plane. To obtain magnetization excitation *eigenmodes* of an equilibrium magnetization state we have applied a short field pulse in the out-of-plane direction with the maximal field value $H_{max} = 100$ Oe and the trapezoidal time dependence with rise and fall times $t_r = t_f = 100$ ps and the plateau duration $t_{pl} = 300$ ps. To obtain the eigenmode spectrum, we have set the dissipation constant to zero ($\lambda = 0$) and recorded magnetization trajectories of each cell during the pulse and for $\Delta t = 10$ ns after the pulse was over. Spatial profiles of the eigenmodes (spatial maps of the oscillation power distribution in the element plane) were then obtained in the meanwhile standard way [2, 22] using the Fourier analysis of these magnetization trajectories after the pulse decay. Because the applied field pulse was spatially homogeneous, we could observe only eigenmodes which symmetry was not lower than the symmetry of the equilibrium magnetization state of the studied system. In principle, the analysis of the eigenvalues and eigenvectors of the energy Hessian matrix [23] allows to obtain all eigenmodes, but our method can be used for much larger systems because it does not require the explicit search of eigenvalues for large matrices with sizes $Z \times Z$ proportional to total number of discretization cells $Z \sim N_x \times N_z \times N_y$. For the qualitative analysis of the influence of various physical factors on the magnetization dynamics aimed in this paper our method provides enough information.

To study the *transient* magnetization dynamics which could be compared to real experiments we have applied the same field pulse as described above and recorded magnetization time dependencies for each discretization cell during the pulse and for $\Delta t = 3$ ns after the pulse. Here the dissipation constant was set to $\lambda = 0.01$ - the value commonly reported in literature for thin Py films; the same constant was used for Co layer. We have checked that increasing this value up to $\lambda = 0.05$ led, as expected, to faster overall oscillation decay, but did not produce any qualitative changes in the magnetization dynamics.

### B. Single layer Py element as the reference system

Keeping in mind that we are going to study interaction effects in a trilayer system, we first present reference results for the single Py nanoelement with the same parameters as the Py layer of the complete trilayer (the square element 1 x 1 mkm$^2$, with the thickness $h_{Py} = 25$ nm, $M_S^{Py} = 860$ G, $A^{Py} = 1 \cdot 10^{-6}$ erg/cm, $K_{cub}^{Py} = 5 \cdot 10^3$ erg/cm$^3$). Fig. 2a shows the equilibrium magnetization structure of such a nanoelement obtained starting from the initial state consisting of four homogeneously magnetized domains in corresponding triangles (as shown, e.g., in Fig. 4a), whereby magnetic moments of four central cells were oriented perpendicular to the layer plane (along the *y*-axis). As expected, the very small random grain anisotropy of Permalloy has virtually no influence on the magnetization state, so that the equilibrium magnetization forms a nearly perfect closed Landau magnetization pattern with four 90° Neel domain walls between the domains and the central vortex showing upwards.

Spectrum of magnetization excitations for this Landau pattern is shown in Fig. 2b together with spatial maps of the oscillation power of the out-of-plane magnetization component for each significant spectral peak. We remind (see Sec. IIA) that the field pulse used to draw the magnetization out of its equilibrium state was spatially homogeneous, and hence only modes with the corresponding symmetry could be excited. Excitation modes of the Landau domain pattern have been recently studied in detail in [10, 11], so here we will only briefly mention several issues important for further comparison with the trilayer system.

The lowest peak in the excitation spectrum in Fig. 2b corresponds, in a qualitative agreement with the results from [10], to the domain wall oscillations, whereby due to the spatial symmetry of the exciting field pulse we observe only the oscillation mode where all domain walls oscillate in-phase. We are not aware of any analytical theory which would allow to calculate the frequency of a 90° domain wall oscillations and thus could be compared to our simulations. From the qualitative point of view, DW oscillations are exchange-dominated and their frequency is the lowest one among other exchange-dominated magnetization excitations, because the equilibrium magnetization configuration inside a DW is inhomogeneous and thus its stiffness with respect to small deviations from the equilibrium is smaller than for a collinear magnetization state.

Peaks with higher frequencies correspond to the oscillations within four triangular domains of the Landau structure; again, only symmetric in-phase oscillations have been observed. According to the analysis performed in [10, 11] domain excitations can be classified into the following types. First, there exist modes which power distribution has nodes (between the peaks) in the radial direction, i.e., from the square center to its edges. Corresponding wave vector is perpendicular to the magnetization direction in the domains ($\mathbf{k} \perp \mathbf{M}$). Such modes are similar to Damon-Eshbach modes in extended thin films and are called radial (wave vector in the radial direction) [10] or transverse (because $\mathbf{k} \perp \mathbf{M}$) [11]. Second, there exist modes with power distribution nodes along the contour around the square center. In this case regions with high power form elongated bands from the center to the edges of the square. For these modes the wave vector of their spatial power distribution is roughly parallel to the local magnetization direction in each domain ($\mathbf{k} \parallel \mathbf{M}$); they behaviour is similar to the backward volume modes in extended thin films. For obvious reasons this second type is called azimuthal [10] or longitudinal [11] modes.

As it can be seen from Fig. 2b, our field pulse excites mainly an azimuthal mode with the frequency $f \approx 3.2$ GHz and several modes which can be classified as mixed radial-azimuthal modes, because their spatial power distribution has nodes along both the radial direction and the contours around the square center. These our results can be compared to simulations from [11], where the Py element with the same lateral sizes 1000 x 1000 nm$^2$, but with the smaller thickness $h = 16$ nm was studied. Qualitatively our power maps are very similar to those shown in [11], but there are some important discrepancies. First of all, our overall excitation spectrum is very different from that presented in [11] (compare our Fig. 2b with Fig. 1d from [11]), although several peak positions are very close. Our power maps for specific modes also have some qualitative similarities to several maps presented in [11], but detailed comparison does not make much sense due to the different total power spectra as mentioned above. All these difference may arise because the simulated nanoelement in [11] was not discretized in the layer plane, but we believe that the major reasons are (*i*) the much shorter excitation pulse ($t_d$ = 2.5 ps pulse length) used in [11] compared to our (300 ps) and (*ii*) the presence of the finite damping in simulations from [11]. This problem requires further investigation, but is beyond the scope of this paper.

Transient magnetization dynamics for the single-layer Py element after the application of the same field pulse as used for the studies of the excitation spectrum is shown in Fig. 3. We remind that for these simulations we have used the non-zero damping $\lambda = 0.01$ typical for Py films. Fig 3 shows the time dependence of the angle between the average layer magnetization and the element plane $\Psi(t) \sim m_\perp(t)$ (panel (a)), spatial maps of the out-of-plane magnetization projection during the pulse (b) and after the pulse (c). By displaying the out-of-plane magnetization projection we have subtracted the equilibrium magnetization $\mathbf{m}^{eq}(\mathbf{r})$, so that maps in Fig. 3 (and all other figures where the transient magnetization dynamics is shown) represent the difference $\Delta m_\perp = m_\perp(\mathbf{r},t) - m_\perp^{eq}(\mathbf{r})$. Homogeneous grey background around the magnetic element shows the reference grey intensity for $\Delta m_\perp = 0$.

First of all, we emphasize that even the small damping $\lambda = 0.01$ used here leads to relatively fast oscillation decay (within ~ 3 ns after the pulse). After the initial increase of the perpendicular magnetization projection due to the field pulse (see the bright contrast across the whole square in Fig. 3b) is over, the time dependence of the average magnetization is dominated by relatively fast oscillations of the domains, slightly modulated by oscillations of a lower frequency due to the domain wall motion. Corresponding patterns can be seen in Fig. 3c, where one can directly recognize that domain walls and domains themselves oscillate with very different frequencies. Further, comparison of the time-dependent maps from Fig. 3c with the spatial power maps in Fig. 2b shows the qualitative relation between the eigenmodes and the transient dynamics of the Py square in this case: not only the contrast due to the DW oscilla-tions, but also characteristic wave patterns inside the domains and near the outer regions of domain walls agree qualitatively with the eigenmode power distributions shown in Fig. 2b.

Analogous simulations (with qualitatively similar results) have been carried out in [24] in order to explain the magnetization dynamics observed there using the time-resolved X-ray microscopy. We shall return to the analysis of these results by comparing our simulation with experimental data in Sec. III.

### C. Trilayer Co/Cu/Py element: Landau structures with the same rotation sense in both magnetic layers

*1. Deformation of the quasistatic magnetization structure*

In this section we consider the trilayer element Co(25nm)/Cu(10nm)/Py(25nm), 1 x 1 mkm$^2$ in-plane size, with magnetic parameters given in Sec. II.B and *cubic* random anisotropy of Co grains with $K_{cub}^{Co} = 6 \cdot 10^5$ erg/cm$^3$. In order to determine the equilibrium magnetization state of any system by minimizing its magnetic free energy we have to choose the initial (starting) magnetization state. As such a state we take in this section for both Co and Py layers the closed in-plane magnetization configuration with sharply formed four triangular domains and four magnetic moments in the middle of each layer pointing in the same out-of-plane direction - along the +*y*-axis (to fix the orientation of the central vortex). An important point is that the rotation sense of the starting magnetization state is *the same* for both layers. This initial state is shown schematically in Fig. 4a. The situation, when the initial state consists of two closed magnetization configurations with *opposite* rotation senses in Co and Py layers, is considered in the next subsection.

The corresponding equilibrium state which comes out as the result of the energy minimization is shown in Fig. 4c. The most striking feature of this state is the strong deformation of a 'normal' Landau pattern (see Fig. 2a). Namely, central vortices in Co and Py layers are significantly displaced in opposite directions and domain walls are bended - also in opposite directions for Co and Py. Equilibrium domains in both layers do not have anymore a shape of isosceles triangles, but rather form 'triangles' with slightly bended sides of different lengths. The degree of the deformation described above depends both on the Cu spacer thickness and the lateral size of the squared trilayer structure (results not shown).

The reason for this unusual deformation can be understood by analyzing the intermediate magnetization configurations arising during the energy minimization. At the first stage of this process the 'normal' Landau domain configuration inside each layer is formed. It is well known that the 90$^o$ Neel domain walls of this magnetization configuration possess both volume and surface 'magnetic charges' along them. We consider in detail the configuration of surface charges, because the density of these 'charges' is simply proportional to the out-of-plane magnetization component $m_\perp(\mathbf{r})$ and is thus easier to visualize. The corresponding map of the out-of-plane magnetization $m_\perp(\mathbf{r})$ for the equilibrium Landau state of a 25 nm thick square Py nanoelement is shown in Fig. 4b. One can clearly see significant enhancement of the $m_\perp(\mathbf{r})$-magnitude along all four domain walls. Thus two lines of the opposite 'surface magnetic' charges are formed along each wall, building a kind of a 'linear dipole' (shown schematically in Fig. 4b with arrows and + and – signs). Now, it is important to realize that the orientation of these dipoles for the given domain wall is the same on *both upper* and *lower* surfaces of the nanoelement. For this reason, for the geometry shown in Fig. 1 and initial magnetization states with the same rotation senses (as shown in Fig. 4a) we have on the *upper*

Co surface and *lower* Py surface linear dipoles with the *same* (parallel) orientations along all four walls in each layer. The volume 'charges' formed due to the non-zero magnetization divergence in the nanoelement volume have a qualitatively similar distribution. They also contribute to the effect described below; with the decrease of the film thickness, when the Neel wall tends to a perfectly 'in-plane' magnetization structure, the contribution from these volume charges becomes dominating.

The linear dipoles described above dipoles obviously repel each other, and due to the small interlayer distance (which is in this case significantly smaller than the wall width) this repulsion is very strong. With other words, corresponding 90° Neel domain walls in Co and Py layers 'feel' a strong mutual repulsion, so that they start to move away from each other. As the result, domain walls in Co and Py layers shift in opposite directions, forming the final equilibrium structure displayed in Fig. 4c. We note in passing that for the starting state used in these simulations, the central vortices in Co and Py layer have the same orientation and thus attract each other. However, due to the small vortex area this attraction can not compensate for the strong repulsion of all domain walls, although the surface density of magnetic charges within the vortices is much higher than along the walls to due a large values of $m_\perp(\mathbf{r})$ within the vortex.

For further consideration it also important to note that the equilibrium magnetization state of Co is disturbed by its random grain anisotropy more than for the Py layer, for which the influence of this anisotropy is very small. Corresponding disturbance can be seen on the spatial map of $m_\perp(\mathbf{r})$ for Co (Fig. 4c, middle panel of the upper row), but for the moderate cubic anisotropy of Co $K_{\text{cub}}^{\text{Co}} = 6 \cdot 10^5$ erg/cm$^3$ and the small average crystallite size $\langle D \rangle = 10$ nm, this disturbance is still rather weak.

*2. Eigenmodes and transient magnetization dynamics*

Strong deformation of the equilibrium domain structure discussed in the previous subsection has a qualitative impact on magnetization dynamics in the trilayer as compared to a single-layer systems.

First of all, spectrum of eigenmodes of the Py layer from Co/Cu/Py trilayer (shown in Fig. 5, upper panel) is qualitatively different from the corresponding single-layer Py square (Fig. 2b). Irregular domain structure results in a quasi-continuos (for our resolution) oscillation power spectrum, because peaks corresponding to the oscillations of each domain and domain wall are located at different positions. In particular, all domain walls oscillate with various frequencies, as shown in Fig. 5 by the peaks of the 1$^{\text{st}}$ group (a, b, c). These peaks can be attributed to oscillations of different domain walls as displayed on the power spatial maps in the upper row of these maps. Eigenmode frequencies for oscillations within the domains also differ significantly for different Py domains, as shown by the peak positions of the 2$^{\text{nd}}$ group in the spectrum. In addition, the power distribution patterns within each domain become highly irregular, as shown by corresponding maps in the second map row in the same figure. For higher frequencies (group 3), the power distribution is even more complicated, although some typical attributes of longitudinal and transverse modes can still be recognized (3$^{\text{rd}}$ map row).

Transient magnetization dynamics of the same trilayer system with the finite damping (it was set to $\lambda = 0.01$ for both Co and Py layers) also strongly differs from the monolayer case. Corresponding time dependencies for the out-of-plane angles $\Psi(t) \sim m_\perp(t)$ of the average magne-tization are shown in Fig. 6a for the Co layer (thin solid line), Py layer (dashed line) and the total system (thick solid line).

The out-of-plane magnetization deviation during the field pulse is smaller for the Co layer than for the Py layer, due to the higher saturation magnetization of Co which leads to larger demagnetizing field caused by the out-of-plane excursion of Co magnetization. Due to the same larger saturation magnetization of Co (and equal Co and Py layer thicknesses), the basic oscillation frequency is now close to the oscillation frequency of domains for the single-layer Co nanosquare: Co layer in the trilayer element determines the overall oscillation frequency, 'locking' (capturing) the frequency of the Py layer domains also. Because the eigenfrequencies of Co and Py domains do not coincide, this phenomenon leads to a much faster decay of the domain oscillations in the Py layer compared to the case of the single-layer Py element (compare the dashed lines in Fig. 6a to Fig. 3a). In fact, shortly after the pulse is over ($t > 0.6$ ns), Py domain oscillations are barely visible both in the average magnetization time-dependence (Fig. 6a) and spatial maps of the out-of-plane magnetization component (Fig. 6c). Low-frequency oscillations of the average magnetization of Py are entirely determined by the oscillations of bended DWs. Note that oscillations of different walls are out of phase due to the different eigenfrequencies of the four DWs in the disturbed Landau structure (see Fig. 5), so that for the given time moment different walls (and even different regions of one and the same wall) can exhibit opposite magnetization contrasts as displayed on the last images in both map rows in Fig. 6b and 6c. We shall return to this important circumstance by comparing our simulations to experimental data in Sec. III.

### D. Trilayer Co/Cu/Py element: Landau structures with opposite rotation senses in Co and Py magnetic layers

*Equilibrium magnetization configuration*. It is well known that the initial magnetization state used to start the energy minimization in micromagnetics can have a decisive influence on the equilibrium configuration resulting from this minimization, because any realistic ferromagnetic system possesses many energy minima due to several competing interactions present in ferromagnets. For this reason we have studied the influence of the starting configuration on the equilibrium magnetization and dynamical properties of our trilayer system, choosing as an alternative starting state the same Landau-like domains structure as described at the beginning of subsection *II.C.1*, but with opposite rotation senses for Co and Py layers (see Fig. 7a).

In this case Landau patterns are also formed at the initial energy minimization stage in both magnetic layers. However, closed magnetization states in Co and Py layers have now opposite rotation senses. For this reason linear magnetic dipoles appearing along each domain wall at the upper Co and lower Py surfaces as described in subsection *II.C.1* above, are oriented *antiparallel*. Hence the domain walls (which form these dipoles) attract each other, so that these walls become wider and do not move across the layers. This naturally leads to a nearly symmetrical magnetization configurations in both Co and Py. This configuration is qualitatively similar to a 'normal' Landau pattern in a single square nanoelement, but domain walls are much broader (compare Fig. 7b to Fig. 2a).

Table 1. Energies of equilibrium magnetization states shown in Fig. 4c and 7b.

| Initial magn. state | Total energy (nanoerg) | Anisotropy energy | Exchange energy | Demag. energy |
|---|---|---|---|---|
| The same rotation senses in Co and Py | 4.193 | 2.851 | 0.715 | 0.627 |
| Opposite rotation senses in Co and Py | 3.602 | 2.838 | 0.510 | 0.254 |

It is instructive to compare the energies of equilibrium magnetization configurations obtained from the two different starting states as explained above. From Table 1 one can see that the total energy of the configuration with opposite rotation senses of closed magnetization states in Co and Py layers (Fig. 7b) is *lower* than the energy of the state with the same rotation senses in both layers (Fig. 4c). The total energy decrease is mainly due to the smaller exchange energy (wider domain walls) and demagnetizing energy (attraction of domain wall dipoles) in the 'opposite' state. However, due to the dominant contribution of the magnetocrystalline anisotropy energy (which is nearly equal in both cases) the total energy difference is not very large, so in experimental realizations both states can be expected.

*Magnetization dynamics*: *eigenmodes*. The almost symmetrical equilibrium magnetization state results in the excitation spectrum with much narrower peaks than for the strongly disturbed asymmetrical state considered in the previous subsection. Corresponding spectrum of eigenmodes for the Py layer (from the Co/Cu/Py trilayer) is shown in Fig. 8 together with spatial maps of the oscillation power. All domain walls have now nearly the same oscillation frequency (similar to the situation for the single-layer Py element), but due to the increased width of the domain walls corresponding oscillation regions are also much wider - compare the $1^{st}$ map on Fig. 8 with the $1^{st}$ map on Fig. 2b. Broadening of domain walls manifests itself also in the significant decrease of the corresponding oscillation frequency ($\approx$ 2.2 GHz for the Py layer within the trilayer vs $\approx$ 3.2 GHz for the single-layer Py element).

Spectral peaks corresponding to the oscillations of domains themselves are also much narrower than for the highly asymmetrical state discussed above, so that several modes can be well resolved (Fig. 8). Although the oscillation frequencies for various domains coincides (within our resolution $\Delta f \sim$ 0.1 GHz) and oscillation power patterns for various domains are very similar (at least for modes 2 and 3 shown in Fig. 8), the absolute values of the spatial power significantly differs from domain to domain. We attribute this effect to the random anisotropy fluctuations of the Co layer. It is well known that due to the small average crystallite size these fluctuations are largely 'averaged out' [25]. However, remaining small fluctuations of the out-of-plane magnetization in the Co layer on a large spatial scale have a significant influence on the Py layer eigenmodes due to the large saturation magnetization of Co and small spacer thickness. In particular, these fluctuations may lead to the redistribution of the oscillation power between the domains as can be seen on the spatial power maps in Fig. 8.

*Magnetization dynamics*: *transient behaviour*. Due to the qualitatively different equilibrium magnetization state for the trilayer with oppositely oriented Landau structures in Co and Py layers its transient magnetization oscillations (after the field pulse) for the finite damping case are also very different from both the single-layer square and the trilayer possessing Landau structures with the same rotation senses in both Co and Py layers. Corresponding simulation results are shown in Fig. 9 in the same format as in Fig. 6.

First of all, due to the largely restored symmetry of the equilibrium magnetization configuration, magnetization oscillations of different domain walls and different domains are now 'in-phase'. Due the much 'softer' magnetization configurations of the domains their oscillations have now a much higher amplitude than for the trilayer with 'parallel' Landau structures (compare after-pulse oscillations and magnetization maps in Fig. 6 and Fig. 9). For this reason the relative contribution of domain *wall* oscillations to the time dependence of the average magnetization is almost negligible. It can be seen that domain oscillations are dominated by the propagating spin wave which is excited at the square center (core of the Landau structure). Its wavefront has initially a nearly squared form, but when the wave propagates towards the element edges, its front becomes circular. Several nodes appear along this wave front for the sufficiently long propagation time (see several last maps in Fig. 9) in accordance with the spatial power distribution of the system eigenmodes. However, the propagating time shown in Fig. 9 is too short to establish the nodal structure with as many nodes as shown in Fig. 6.

### E. Trilayer Co/Cu/Py element: influence of the Co anisotropy type

It is well known that thin polycrystalline Co films may possess two kinds of the magnetocrystalline grain anisotropy, according to the two possible grain types: *fcc* grains have the cubic anisotropy $K_{\text{cub}}^{\text{Co}} = 6 \cdot 10^5$ erg/cm$^3$ (the case which was analyzed above) and *hcp* grains have the much stronger uniaxial anisotropy $K_{\text{un}}^{\text{Co}} = 4 \cdot 10^6$ erg/cm$^3$ (see [19] and original experiments in [26] for the corresponding discussion). Films with mixed *fcc-hcp* structure are also possible. For this reasons we have studied the effect of the Co anisotropy type on the equilibrium magnetization structure and magnetization dynamics of our trilayer simulating the system with all parameters as given above, and with the starting Landau states with the same rotation sense in both magnetic layers, but with the uniaxial anisotropy of Co grains $K_{\text{un}}^{\text{Co}} = 4 \cdot 10^6$ erg/cm$^3$. Grain anisotropy axes were again distributed randomly in 3D. Corresponding results are presented in Fig. 10 and 11.

The major effect of such a large magnetocrystalline grain anisotropy is the strong disturbance of the equilibrium magnetization structure, as it can be seen from Fig. 10. Despite the small average grain size $\langle D \rangle = 10$ nm, the anisotropy fluctuations in Co even after their 'averaging-out' [25] are sufficiently strong to induce large deviations from the ideal Landau structure and to enforce significant randomly varying out-of-plane magnetization component on the lateral Co surfaces. The stray field induced on the Py layer by this $m_\perp^{\text{Co}}(\mathbf{r})$-component nearly destroys the original Landau magnetization structure of this layer, so that only the overall magnetization rotation sense is preserved. The initially triangular domains of the Landau structure now have a highly irregular form and only small pieces of domain walls (mainly near the Py square corners) can be recognized (Fig. 10b).

Correspondingly, the magnetization dynamics of such a trilayer element again differs qualitatively from all cases studied above (Fig. 11). Oscillations of the domain walls are almost invisible. Average magnetization time dependence is entirely dominated by the circular wave emitted from the central vortex as shown in Fig. 11b and 11c. Due to the strongly disturbed domain structure and absence of well defined domain walls (at least in the middle of the Py square) the wave front is roughly circular from the very beginning. However, irregularities of the equilibrium magnetization structure within the domains lead to large modulations of the oscillation amplitude along the wave front, as it can be recognized already for the initial stage of the wave propagation (Fig. 11b).

# III. VERIFICATION OF SIMULATION PREDICTIONS AND COMPARISON WITH EXPERIMENTAL OBSERVATIONS

Although, as already mentioned in the Introduction, both static magnetization structures and magnetization dynamics in multilayer nanoelements have been extensively studied in the last several years, we are not aware of any experiments which could be used for direct confirmation or disprove of our simulation results.

In principle, our predictions concerning the *equilibrium* magnetization structure in square mkm-sized multilayer elements (Fig. 4, 8, 10, 12) can be verified quite easily. Fabrication of mkm- and sub-mkm patterned multilayer elements of corresponding sizes is possible using several experimental techniques. Deformation of the 'normal' Landau structure predicted by us is strong enough for both the same and opposite magnetization rotation senses in individual layers of the magnetic element. Hence it should be possible to detect this deformation with the state-of-the-art methods for the observation of magnetization structures in nm-thick layers, e.g., using the meanwhile standard high-resolution MFM-facilities. We believe, that in an array of square nanoelements composed as described in this paper both types of the equilibrium magnetization states (depending on random initial fluctuations of the magnetization) will be formed, so that structures shown both in Fig. 4 and Fig. 7 can be found.

Magnetization *dynamics* of thin film systems can be measured nowadays not only with a very high lateral and temporal resolution, but also element-specific using the synchrotron X-ray radiation [24, 27, 28] with the potential resolution of several tens of nanometers. Layer-selective measurements are also possible using the Kerr microscopy technique [29], whereby the resolution lies in the sub-mkm region (see, e.g., the recent detailed study of the magnetization dynamics of the Landau state for Py squares with sizes ~ 10 - 40 mkm in [32]). As already mentioned in the Introduction, most papers on this topic are devoted to the magnetization dynamics of mkm-sized *single-layer* elements. Excitations for trilayered *circular* Py/Cu/Py nanodots at different external fields were studied in thermodynamical equilibrium in [30]; mainly the interlayer interaction effects due to the magnetic poles on the *edges* of nearly saturated layers have been described. There are also a few papers where the magnetization *switching* of rectangular magnetic trilayers is studied (see, e.g., [31]), where the major effect is also due to the strong stray fields induced near the *edges* of a nanoelement in a magnetically saturated state.

We are aware of only two experimental studies which results can be more or less directly related to the subject of this paper, namely, the interlayer dipolar interaction dominated by the nearly in-plane domain walls of the closed (Landau-like) magnetization configuration. In both cases [24, 27] the magnetization dynamics of Co/Cu/Py square trilayers was studied using the element-specific time-dependent synchrotron X-ray microscopy.

In the pioneering paper [24] the transient magnetization dynamics of a relatively large 4 x 4 mkm$^2$ trilayer Co(50nm)/Cu(2nm)/Py(50nm) was studied by the pump-and-probe X-ray magnetic circular dichroism (XMCD) microscopy in the field pulse perpendicular to the sample plane. This technique has allowed to investigate the magnetization dynamics with the temporal resolution ~ 50 ps and potential spatial resolution ~ 20 nm (however, the actual resolution achieved in [24] is much poorer and is difficult to estimate due to a significant shot noise). Stoll et al. [24] did not study the equilibrium magnetization state of their system and have presented spatial maps of the out-of-plane magnetization component of the Py layer only for various time moments during and after the field pulse. Magnetization maps shown in [24] are effectively differential images between the excited and equilibrium magnetic states, so that the contrasts due to the *static* domain walls and central vortex of the closed magnetization structure are excluded.

The main qualitative features of experimental images presented in [24] are the following: (*i*) bright and relatively narrow bands along the diagonals of the squared magnetic element, i.e., where the domain walls of the 'normal' equilibrium Landau pattern are located; (*ii*) at the same time different domain walls exhibit contrast of *different brightness* and even of *different signs*, indicating that oscillations phases and/or frequencies for different walls are different; (*iii*) after the decay of the field pulse virtually *no contrast within the domains* themselves can be seen, so that the average out-of-plane magnetization component after the field pulse is zero ($\langle m_{\perp}^{Py}(\mathbf{r})\rangle = 0$) within the experimental resolution.

The authors of [24] attributed the narrow contrast bands mentioned above to the domain walls oscillations. To support their experimental findings, Stoll et al. have performed dynamic micromagnetic simulations, where they have included the *Py layer only* and discretized this layer only in the lateral plane. Simulated out-of-plane magnetization images shown in [24] demonstrate, of course, the time-dependent contrast between the oscillations of domain walls and domains themselves, but clearly fail to reproduce all other qualitative features of their experimental images listed above.

We have also performed simulations of the Py single layer element with the sizes used in [24], discretizing it into 400 x 400 x 4 (totally 6.4·10$^5$) cells. We note that due to the low anisotropy and relatively low saturation magnetization of Py the size of our discretization cells 10 x 10 x 12.5 nm$^3$ was small enough to reproduce main features of the Py dynamics. Proper simulation of the magnetic trilayer with the same lateral sizes including the 50 nm thick Co layer would require to halve the cell size in each dimension, so that the overall cell number would be prohibitively large for the state-of-the-art micromagnetic simulations. Our $m_{\perp}^{Py}(\mathbf{r})$-images (Fig. 12) qualitatively agree with simulation data from [24], demonstrating once more that when the interaction with the Co layer is neglected, *in-phase* oscillations of all four DWs of the Landau pattern should be observed. In addition, the strong contrast within the domains after the field pulse is clearly seen in Fig. 12c, manifesting itself also in strong after-pulse oscillations of the *average* out-of-plane magnetization $\langle m_{\perp}^{Py}(t)\rangle$ as shown in Fig. 12a. The amplitude of these after-pulse oscillations is comparable with the maximal value of $\langle m_{\perp}^{Py}\rangle$ achieved during the pulse. Taking into account that the domain contrast during the pulse is clearly seen in the experimental images presented in [24], it is unlikely that approximately the same contrast after the pulse would be completely overlooked. All in one, experimental findings from [24] can not be explained satisfactory when the dynamics of Co layer and the interlayer interaction in the trilayer Co/Cu/Py is neglected.

Taking into account that we could not simulate (at least not with proper resolution) the complete system studied in [24], we can compare the results of Stoll et al. with our simulation data only qualitatively. First of all, we note that the straight lines corresponding to the domain wall oscillations indicate that Co and Py layers in this experiment possess Landau magnetization states with opposite rotation senses, because for the trilayer with the same magnetization rotation senses in both magnetic layers domain walls should be strongly bended (see Fig. 4 and 6 above).

From the remaining possibilities, dynamic magnetization images of the trilayer with 'opposite' Landau patterns and *fcc* Co crystallites (Fig. 9) demonstrate well pronounced straight lines corresponding to the domain wall oscillations similar to those observed in [24]. However, due to the high symmetry of the equilibrium state, all domain walls oscillate in-phase and with the same amplitude, in contrast with strongly out-of-phase DW-oscillations with different amplitudes seen in Fig. 2 from [24]. For the same trilayer with *hcp*-Co magnetization dynamics images are asymmetric (Fig. 11), but due to the very blurred boundaries between Py domains virtually no contrast is observed along the square diagonals (which would correspond to DW-oscillations).

At this point it should be noted that the thickness of magnetic layers studied in [24] ($h_{Co} = h_{Py}$ = 50 nm) is twice as large as in our simulated system ($h_{Co} = h_{Py}$ = 25 nm). In a system with such thick layers, domain wall structure disturbed due to the interlayer interaction, could be partially recovered due to thicker magnetic layers. To check this idea, we have simulated the Co/Cu/Py trilayer with the same parameters and initial magnetization structure, as for the system shown in Fig. 10 and 11, but with the Py thickness $h_{Py}$ = 50 nm; here the Py layer was discretized in 8 in-plane sublayers, so that the size of the discretization cell was preserved. Corresponding simulation results are shown in Fig. 13 (equilibrium state) and 14 (magnetization dynamics). One can see, that for such increased Py thickness domain walls in the equilibrium magnetization state are, indeed, partially recovered (see Fig. 13), so that their oscillations are clearly visible in the dynamic patterns (Fig. 14). Due to the remaining asymmetry of the magnetization structure, oscillations of different DWs have different spatial patterns, amplitudes and frequencies, in a qualitative agreement with the images displayed in [24]. The average out-of-plane magnetization projection exhibits only very weak oscillations after the field pulse, which is also in agreement with [24]. However, we observe significant magnetization contrast near the square center which is due to the wave emitted by the vortex core; this contrast was not found experimentally [24].

In the second paper mentioned above [27] the magnetization dynamics of 1 x 1 mkm$^2$ Co(20nm)/Cu(10nm)/Py(20nm) trilayer was studied in the *in-plane* pulsed magnetic field, so that mainly the central vortex motion could be seen both in Co and Py layers. The authors display also the equilibrium magnetization structures of both magnetic layers (see Fig. 1 in [27]), also obtained using XMCD-microscopy. Unfortunately, the resolution of these images is still not good enough to make any quantitative statements, so one can only say that both Co and Py layers possess closed magnetization structures with the same rotation senses and that these structures are somewhat disturbed (compared to 'ideal' Landau patterns). However, no meaningful quantitative comparison to our results presented in Fig. 4 is possible.

### IV. CONCLUSION

In this paper we have studied the effects of magnetodipolar interlayer interaction in trilayer elements with lateral sizes in sub-mkm region and magnetic layers and spacer thicknesses of several nanometers. We have shown that due to such a small interlayer distance even relatively weak stray field induced by the 90° Neel domain walls of the *closed magnetization state* (Landau-like pattern) causes qualitative changes of both the *equilibrium* magnetization structure and magnetization *dynamics* in these systems. We have also demonstrated that the effect of such an interaction may be very different, depending not only on the initial magnetization state used to find the equilibrium magnetization pattern of a system, but also on the crystallographic structure of magnetic layers. This random crystal grain structure significantly affects the magnetization dynamics also for very small crystallite size, where the random magnetocrystalline anisotropy of the grains is largely 'averaged-out'. The statement about this anisotropy averaging is often used to justify the neglect of this random anisotropy when simulating the corresponding magnetization dynamics; our results reveal that in many important cases such a neglect may be the prohibitive oversimplification of a problem.

Our simulations clearly demonstrate that for the qualitative and especially quantitative understanding of magnetization dynamics in multilayers, magnetodipolar interlayer interaction effects must be included into consideration, even when the equilibrium magnetization structure forms a closed flux state and thus its stray field is believed to be relatively weak.

Although we are not aware of any experimental studies which results could be directly compared to our simulation data, our main predictions can be relatively easily verified with available experimental techniques, as discussed in detail in Sec. III.. However, the accurate sample characterization both from crystallographic and magnetic points of view is required to enable a meaningful comparison with experimental results.

**ACKNOWLEDGEMENTS**. The authors acknowledge fruitful discussions with P. Fischer and A. Slavin. This research was partly supported by the Deutsche Forschungsgemeinschaft (DFG grant BE 2464/4-1).

**References:**

[1] U. Hartmann (Ed.), *Magnetic Multilayers and Giant Magnetoresistance: Fundamentals and Industrial Applications, Springer Series in Surface Sciences*, Vol. 37, 2000; H. Hopster and H.P. Oepen (Eds.), *Magnetic Microscopy of Nanostructures, Series: NanoScience and Technology*, 2005; H. Zabel and S.D. Bader (Eds.), *Magnetic Heterostructures: Advances and Perspectives in Spinstructures and Spintransport*, *Springer Tracts in Modern Physics*, Vol. 227, 2007

[2] C. Bayer, J. Jorzick, B. Hillebrands, S.O. Demokritov, R. Kouba, R. Bozinoski, A.N. Slavin, K. Guslienko, D.V. Berkov, N.L. Gorn, M. P. Kostylev, Spin-wave excitations in finite rectangular elements of $Ni_{80}Fe_{20}$, *Phys. Rev. B*, **72** (2005) 064427

[3] R. D. McMichael, M. D. Stiles, Magnetic normal modes of nano-elements, *J. Appl. Phys.*, **97** (2005) 10J901

[4] V.V. Kruglyak, P.S. Keatley, R.J. Hicken, J.R. Childress, J.A. Katine, Time resolved studies of edge modes in magnetic nanoelements (invited), *J. Appl. Phys.*, **99** (2006) 08F306

[5] J. Podbielski, F. Giesen and D. Grundler, Spin-Wave Interference in Microscopic Rings, *Phys. Rev. Lett.*, **96** (2006) 167207

[6] M. Buess, T. Haug, M.R. Scheinfein, C.H. Back, Micromagnetic Dissipation, Dispersion, and Mode Conversion in Thin Permalloy Platelets, *Phys. Rev. Lett.*, **94** (2005) 127205; G. Gubbiotti, G. Carlotti, T. Okuno, M. Grimsditch, L. Giovannini, F. Montoncello, and F. Nizzoli, Spin dynamics in thin nanometric elliptical Permalloy dots: A Brillouin light scattering investigation as a function of dot eccentricity, *Phys. Rev.*, **B72** (2005) 184419; K. Yu. Guslienko, W. Scholz, R. W. Chantrell and V. Novosad, Vortex-state oscillations in soft magnetic cylindrical dots, *Phys. Rev.*, **B71** (2005) 144407; X. Zhu, Z. Liu, Vitali Metlushko, P. Grütter, M.R. Freeman, Broadband spin dynamics of the magnetic vortex state: Effect of the pulsed field direction, *Phys. Rev. B*, **71** (2005) 180408; G. Gubbiotti, M. Madami, G. Carlotti, H. Tanigawa, T. Ono, L. Giovannini, F. Montoncello, Splitting of Spin Excitations in Nanometric Rings Induced by a Magnetic Field, *Phys. Rev. Lett.*, **97** (2006) 247203; K. Yu. Guslienko, X. F. Han, D. J. Keavney, R. Divan and S. D. Bader, Magnetic Vortex Core Dynamics in Cylindrical Ferromagnetic Dots, *Phys. Rev. Lett.*, **96** (2006) 067205; I. Neudecker, K. Perzlmaier, F. Hoffmann, G. Woltersdorf, M. Buess, D. Weiss, C. H. Back, Modal spectrum of permalloy disks excited by in-plane magnetic fields, *Phys. Rev.*, **B73** (2006) 134426

[7] K. Perzlmaier, M. Buess, C.H. Back, V.E. Demidov, B. Hillebrands, S.O. Demokritov, Spin-Wave Eigenmodes of Permalloy Squares with a Closure Domain Structure, *Phys. Rev. Lett.*, **94** (2005) 057202

[9] J. Raabe, C. Quitmann, C.H. Back, F. Nolting, S. Johnson, C. Buehler, Quantitative Analysis of Magnetic Excitations in Landau Flux-Closure Structures Using Synchrotron-Radiation Microscopy, *Phys. Rev. Lett.*, **94** (2005) 217204

[10] M. Yan, G. Leaf, H. Kaper, R. Camley, M. Grimsditch, Spin-wave modes in a cobalt square vortex: Micromagnetic simulations, *Phys. Rev.*, **B73** (2006) 014425

[11] M. Bolte, G. Meier, C. Bayer, Spin-wave eigenmodes of Landau domain patterns, *Phys. Rev. B*, **73** (2006) 052406

[12] W. Kuch, Magnetic Imaging, in: Lect. Notes Phys., **697** (2006) 275, Springer-Verlag, Berlin Heidelberg

[13] K. S. Buchanan, K.Yu. Guslienko, A. Doran, A. Scholl, S. D. Bader, V. Novosad, Magnetic remanent states and magnetization reversal in patterned trilayer nanodots, *Phys. Rev. B*, **72** (2006) 134415

[14] A. Hubert, R. Schäfer, Magnetic Domains: The Analysis of Magnetic Microstructures, Springer Verlag, Heidelberg, 1998

[15] L. Thomas, M.G. Samant and S. S. P. Parkin, Domain-Wall Induced Coupling between Ferromagnetic Layers, *Phys. Rev. Lett.*, **84** (2000) 1816

[16] W. S. Lew, S. P. Li, L. Lopez-Diaz, D. C. Hatton, and J. A. C. Bland, Mirror Domain Structures Induced by Interlayer Magnetic Wall Coupling, *Phys. Rev. Lett.*, **90** (2003) 217201

[17] V. Christoph, R. Schäfer, Numerical simulation of domain walls in Fe whiskers and their interaction with deposited thin films, *Phys. Rev.*, **B70** (2004) 214419

[18] J. Vogel, W. Kuch, R. Hertel, J. Camarero, K. Fukumoto, F. Romanens, S. Pizzini, M. Bonfim, F. Petroff, A. Fontaine, and J. Kirschner, Influence of domain wall interactions on nanosecond switching in magnetic tunnel junctions, *Phys. Rev.*, **B72** (2005) 220402(R)

[19] D.V. Berkov, N.L. Gorn, Magnetization precession due to a spin-polarized current in a thin nanoelement: Numerical simulation study, *Phys. Rev.,* **B72** (2005) 094401

[20] D.V. Berkov, N.L. Gorn, MicroMagus - package for micromagnetic simulations, http:\\www.micromagus.de

[21] Y. Tserkovnyak, A. Brataas, G.E.W. Bauer, B.I. Halperin, Nonlocal magnetization dynamics in ferromagnetic heterostructures, *Rev. Mod. Phys.*, **77** (2005) 1375

[22] D.V. Berkov, N.L. Gorn, Stochastic dynamic simulations of fast remagnetization processes: recent advances and applications, *J. Magn. Magn. Mat.*, **290-291P1** (2005) 442

[23 ] M. Grimsditch, L. Giovannini, F. Montoncello, F. Nizzoli, G.K. Leaf, H.G. Kaper, *Phys. Rev. B*, **70** (2004) 054409

[24] H. Stoll, A. Puzic, B. van Waeyenberge, P. Fischer, J. Raabe, M. Buess, T. Haug, R. Höllinger, C. Back, D. Weiss, G. Denbeaux, High-resolution imaging of fast magnetization dynamics in magnetic nanostructures, *Appl. Phys. Lett.*, **84** (2004) 3328

[25] G. Herzer, *Nanocrystalline Soft Magnetic Alloys*, in: *Handbook of Magnetic Materials*, vol. 10, Ed. by K. Buschov, Elsevier Science, 1997

[26] D. Weller, G.R. Harp, R.F.C. Farrow, A. Cebollada, J. Sticht, Orientation dependence of the polar Kerr effect in hcp and fcc Co, *Phys. Rev. Lett.*, **72** (1994) 2097; R.M. Osgood III, K.T. Riggs, A.E. Johnson, J.E. Mattson, C.H. Sowers, S.D. Bade, Magneto-optic constants of hcp and fcc Co films, *Phys. Rev.*, **B56** (1997) 2627; K. Heinz, S. Müller, L. Hammer, Crystallography of ultrathin iron, cobalt and nickel films grown epitaxially on copper, *J. Phys.: Cond. Matt.*, **11** (1999) 9437; J. Langer, R. Mattheis, B. Ocker, W. Mass, S. Senz, D. Hesse, J. Kräusslich, Microstructure and magnetic properties of sputtered spin-valve systems, *J. Appl. Phys.*, **90** (2001) 5126

[27] K. W. Chou, A. Puzic, H. Stoll, G. Schütz, B. van Waeyenberge, T. Tyliszczak, K. Rott, G. Reiss, H. Brückl, I. Neudecker, D. Weiss, C. H. Back, Vortex dynamics in coupled ferromagnetic multilayer structures, *J. Appl. Phys.*, **99** (2006) 08F305

[28] Y. Guana, W. E. Bailey, C.-C. Kao, E. Vescovo, D. A. Arena, Comparison of time-resolved x-ray magnetic circular dichroism measurements in reflection and transmission for layer-specific precessional dynamics measurements, *J. Appl. Phys.*, **99** (2006) 08J305

[29] R. Schäfer, R. Urban, D. Ullmann, H. L. Meyerheim, B. Heinrich, L. Schultz, and J. Kirschner, Domain wall induced switching of whisker-based tunnel junctions, *Phys. Rev.*, **B65** (2002) 144405

[30] G. Gubbiotti, M. Madami, S. Tacchi, G. Carlotti, T. Okuno, Field dependence of spin excitations in NiFe/Cu/NiFe trilayered circular dots, *Phys. Rev.*, **B73** (2006) 144430

[31] J. X. Zhang, L. Q. Chen, Magnetic reversal of double-layer patterned nanosquares, *J. Appl. Phys.*, **97** (2005) 084313

[32] A. Neudert, J. McCord, D. Chumakov, R. Schäfer, L. Schultz, Small-amplitude magnetization dynamics in permalloy elements investigated by time-resolved wide-field Kerr microscopy, *Phys. Rev.*, **B71** (2005) 134405

**Figure captions:**

Fig. 1. Geometry of the simulated system, co-ordinate axes and the pulsed field direction used in simulations.

Fig. 2. Magnetic properties of the single-layer squared *Permalloy* element (1000 x 1000 x 25 nm$^3$): (a) - equilibrium Landau magnetization structure in zero external field shown as grey scale maps of magnetization projections; (b) - spectrum of eigenmodes excited by the small out-of-plane pulsed homogeneous field for the state shown in (a). Grey-scale maps below the spectrum show the spatial distribution of the oscillation power for corresponding peaks (bright areas correspond to a large oscillation power)

Fig. 3. Magnetization dynamics of a single-layer *Permalloy* element with the same sizes as in Fig. 2 in the pulsed out-of-plane field: (a) - time dependence of the angle $\Psi_{perp}$ between the average layer magnetization and the element plane ($\Psi_{perp} \sim \langle m_y(\mathbf{r}) \rangle$, see Fig. 1); the trapezoidal pulse form is shown at the same panel as thin solid line; (b) - grey scale maps of $m_y(\mathbf{r})$ (out-of-plane magnetization projection) during the pulse; (c) grey scale maps of $m_y(\mathbf{r})$ after the pulse. Oscillations of both domain walls and domains themselves are clearly seen.

Fig. 4. To the formation of a static equilibrium magnetization structure in zero external field for the trilayer Co/Cu/Py element with the lateral sizes 1000 x 1000 nm$^2$, Co and Py thicknesses $h_{Co} = h_{Py} = 25$ nm and spacer thickness $h_{Cu} = 10$ nm. The Co layer possesses a *fcc* polycrystalline structure with the average crystallite size $\langle D \rangle = 10$ nm and cubic grain anisotropy $K_{cub} = 6 \cdot 10^5$ erg/cm$^3$. (a) - initial magnetization structure used as the starting state by the calculation of the equilibrium structure shown in (c) as grey scale maps of magnetization projections; (b) initial distribution of the surface charges responsible for the repulsion of 90$^o$ domain walls initially located along the main diagonals of the square in Co and Py layers.

Fig. 5. Eigenmodes spectrum for a *Permalloy* layer of the trilayer element with the static magnetization structure shown in Fig. 4c. Due to the symmetry breaking of the underlying magnetization state spectral lines corresponding to the oscillations of different domain walls are positioned at different frequencies (spectral group 1). Spectral peaks corresponding to the domain oscillations form two quasi-continuous groups (groups 2 and 3), whereby each line within a group corresponds to magnetization oscillations within a specific domain as shown by grey-scale maps of the oscillation power distributions below.

Fig. 6. Magnetization dynamics of a trilayer Co/Cu/Py element with the static magnetization structure shown in Fig. 4c in the pulsed out-of-plane field: (a) - time dependence of the angle $\Psi_{perp}$ between the average magnetization and the element plane for the magnetization of the total element (thick green line), Co layer (thin blue line) and Py layer (thick dashed line). Grey scale maps of the out-of-plane magnetization projection of the Py layer during the pulse (b) and after the pulse (c). Due to different eigenfrequencies oscillations of different domain walls are out-of-phase here and oscillation of domains themselves are strongly suppressed compared to the case of a single-layered Py element (Fig. 3c).

Fig. 7. Static equilibrium magnetization structure in zero external field for the same Co/Cu/Py element as shown in Fig. 4, but starting from Landau magnetization states with opposite rotation senses in Co and Py layers (a). Resulting equilibrium state is shown at the panel (b)

as grey scale maps of magnetization projections. It can be seen that due to the attraction of domain walls for the starting magnetization states the symmetry of the final equilibrium magnetization structure is nearly preserved.

Fig. 8. Eigenmodes spectrum for a *Permalloy* layer of the trilayer element with the static magnetization structure shown in Fig. 7b. Due to the largely preserved symmetry of domain walls their oscillations have nearly the same frequency (spectral line 1 and grey-scale map 1). Oscillation power distribution in domain regions (maps 2-4) is still asymmetric due to magnetodipolar interaction with the Co layer which magnetization has a noticeable and spatially varying out-of-plane component due to a significant magnetocrystalline anisotropy (*fcc* Co).

Fig. 9. Magnetization dynamics of a trilayer Co/Cu/Py element with the static magnetization structure shown in Fig. 7b in the pulsed out-of-plane field presented in the same way as in Fig. 6. It can be seen that the magnetization dynamics is largely dominated by the propagation of the spin wave excited at the central vortex; its wave front has initially the square shape which transforms during the propagation into a nearly circular one.

Fig. 10. Static equilibrium magnetization state for $H_{ext} = 0$ for the same Co/Cu/Py element as shown in Fig. 7 (starting from Landau magnetization states with opposite rotation senses in Co and Py layers - see (a)) but with the Co layer having a *hcp* polycrystalline structure with the uniaxial grain anisotropy $K_{un} = 4 \cdot 10^6$ erg/cm$^3$. Due to such a large random anisotropy value the symmetry of the final magnetization state is strongly disturbed (b) and boundary regions between the domains are very wide.

Fig. 11. Magnetization dynamics of a trilayer Co/Cu/Py element with the static magnetization structure shown in Fig. 10b in the pulsed field presented in the same way as in Fig. 9. Due to the strong disturbance of the static magnetization state the oscillations of domain walls are nearly invisible. Although the front of the dominating spin wave remains approximately circular, the wave amplitude shows significant inhomogeneities along this wave front.

Fig. 12. Simulated transient magnetization dynamics for a single-layer Py element with lateral sizes 4 x 4 mkm$^2$ and thickness $h_{Py} = 50$ nm (after the same field pulse and presented in the same way as in Fig. 3). Strong magnetization oscillations within the domain regions after the field pulse can be seen.

Fig. 13. Static equilibrium magnetization state for $H_{ext} = 0$ for the same Co/Cu/Py element as shown in Fig. 10 (starting from Landau states with opposite rotation senses in Co and Py layers - see (a)) but with the thicker Py layer: $h_{Py} = 50$ nm. Due to the increased Py layer thickness domain walls within this layer are partially recovered (see the grey-scale map of $m_y(\mathbf{r})$ for Py in Fig. 13b).

Fig. 14. Magnetization dynamics of a trilayer Co/Cu/Py element with the static magnetization state from Fig. 13b presented in the same way as in Fig. 11. In contrast to the case shown in Fig. 11, oscillations of domain walls can be clearly seen. However, these oscillations remain strongly asymmetric, what can be seen especially well on the magnetization maps after the field pulse (c).

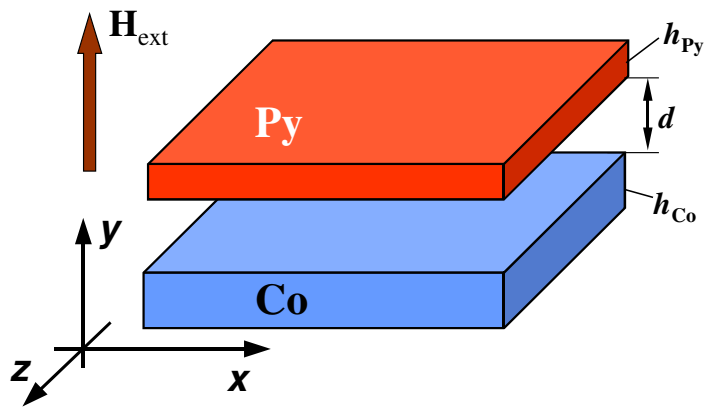

Fig. 1

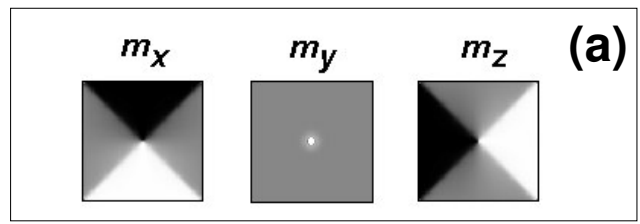

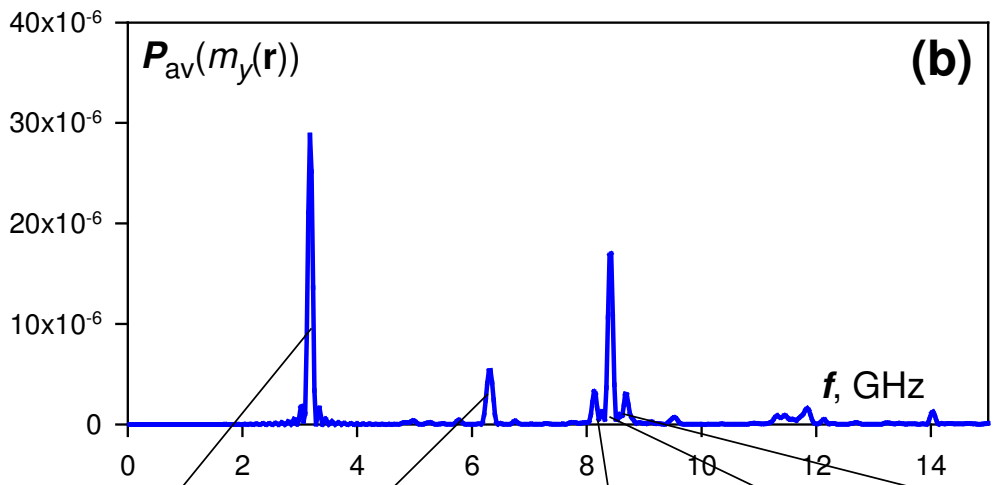

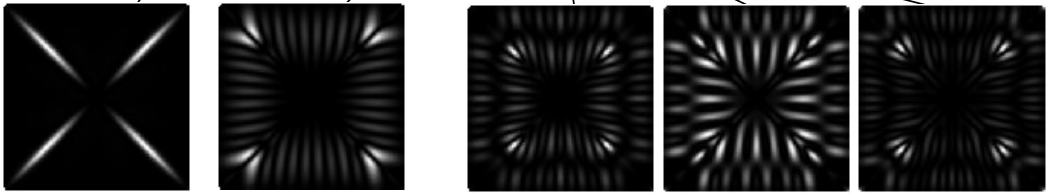

Fig. 2

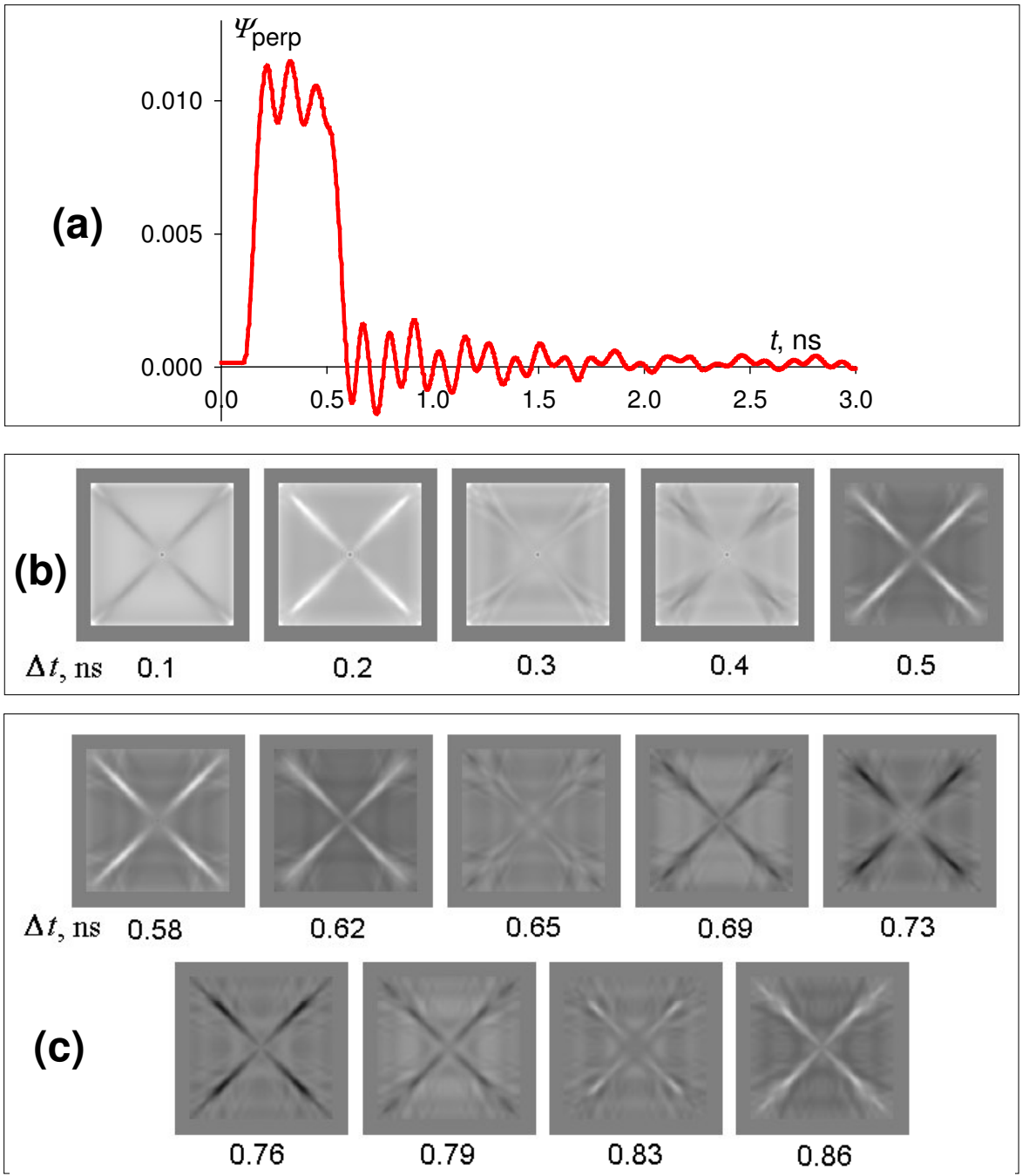

Fig. 3

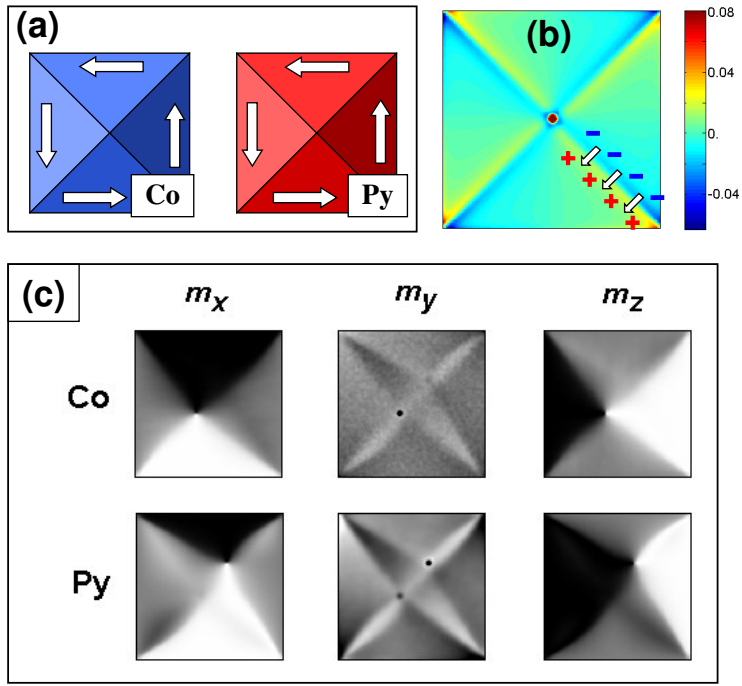

Fig. 4

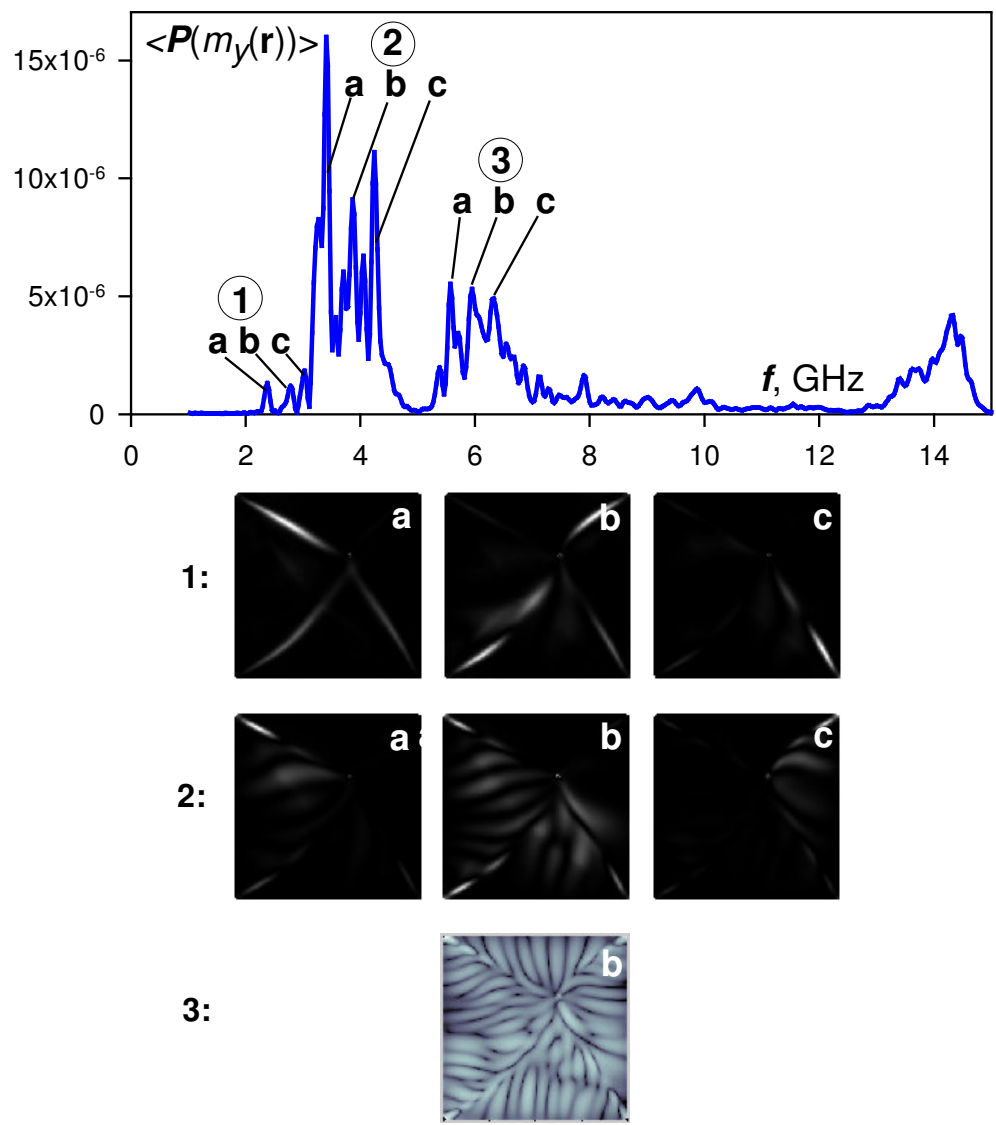

Fig. 5

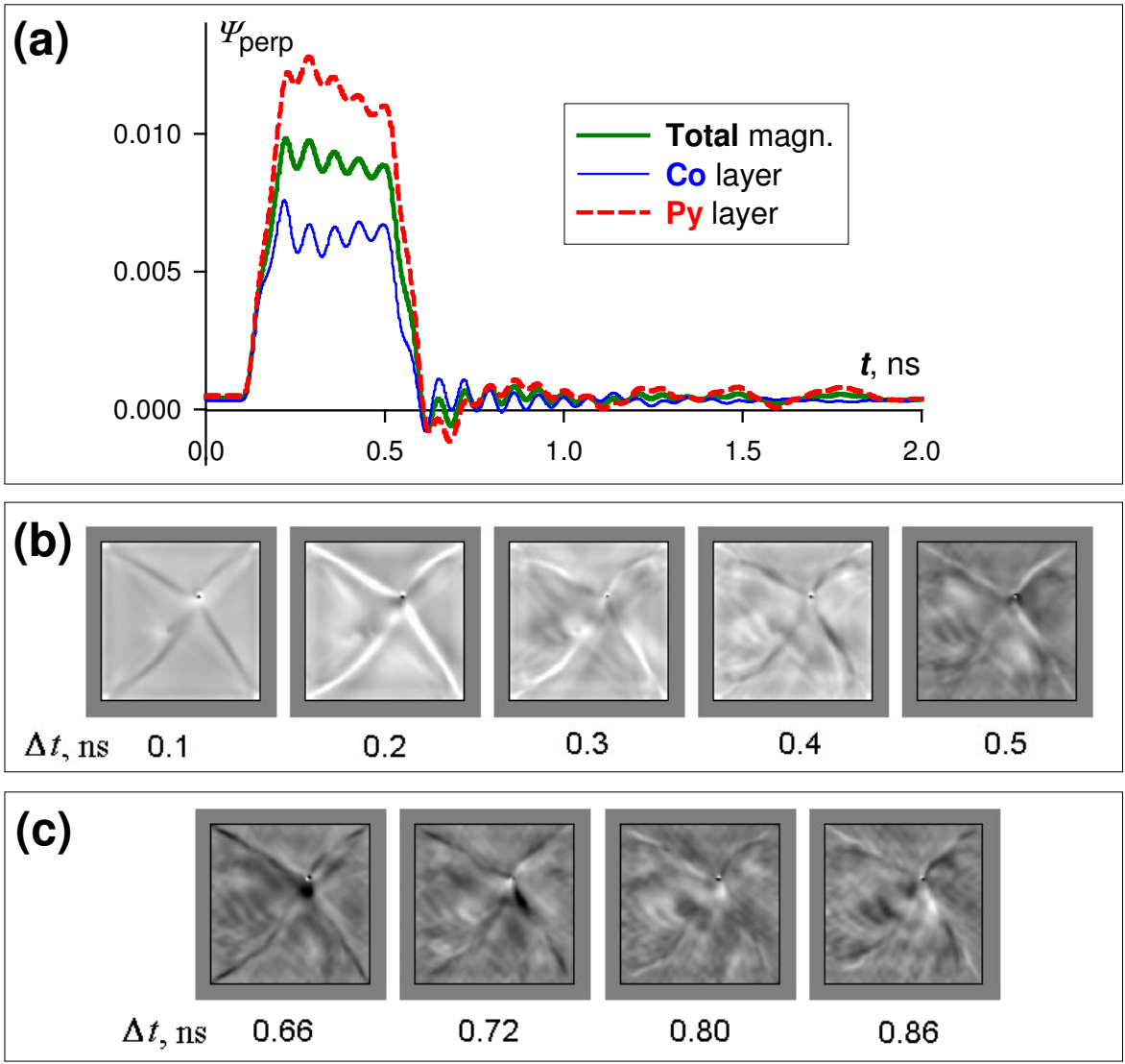

Fig. 6

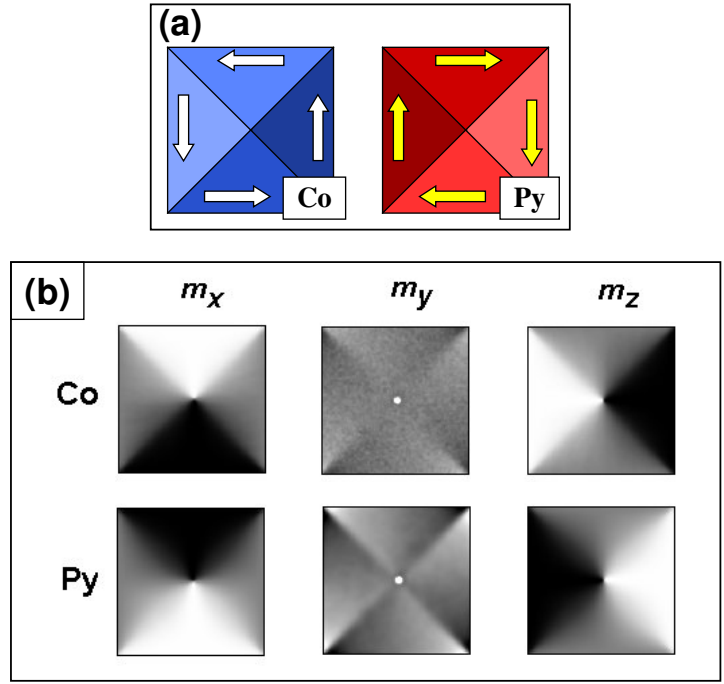

Fig. 7

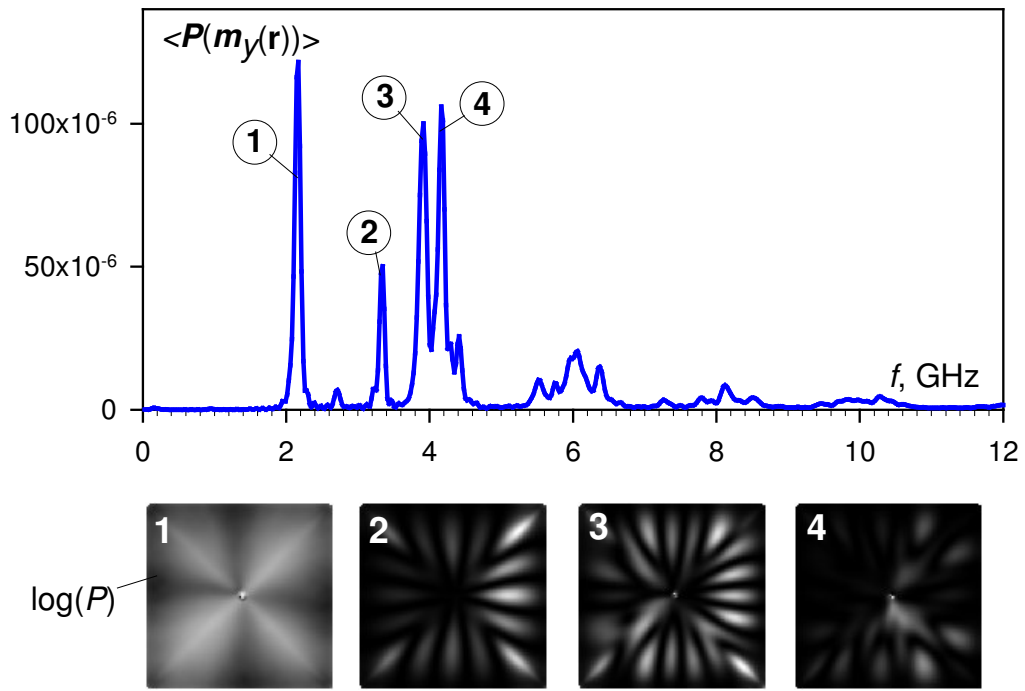

Fig. 8

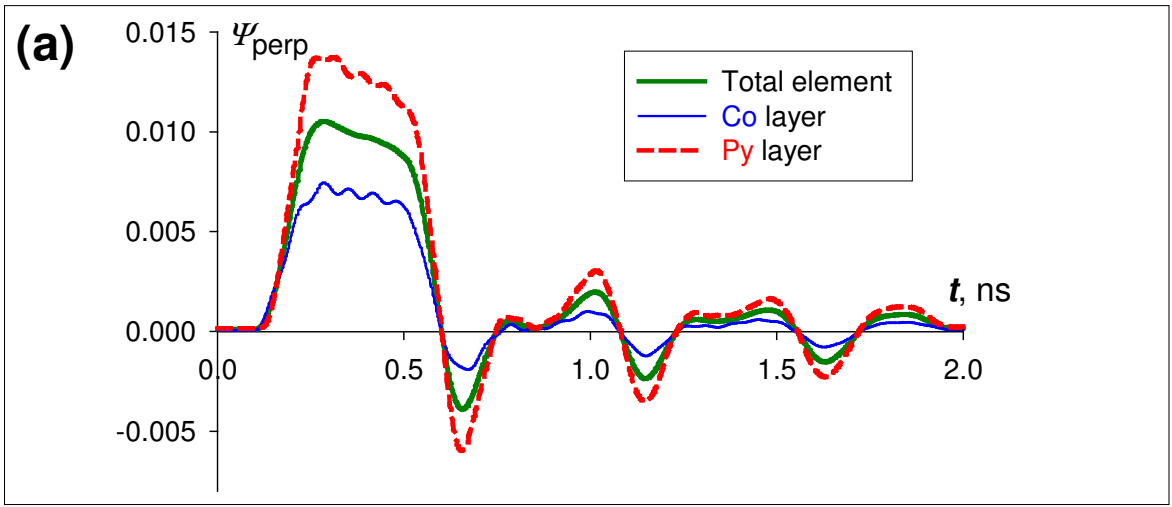

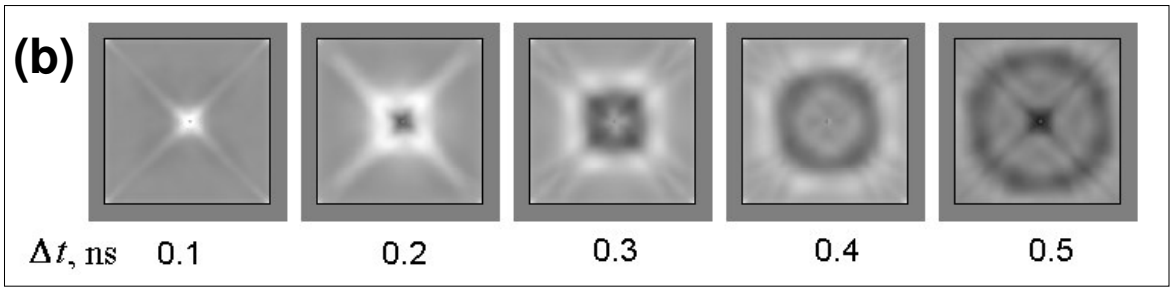

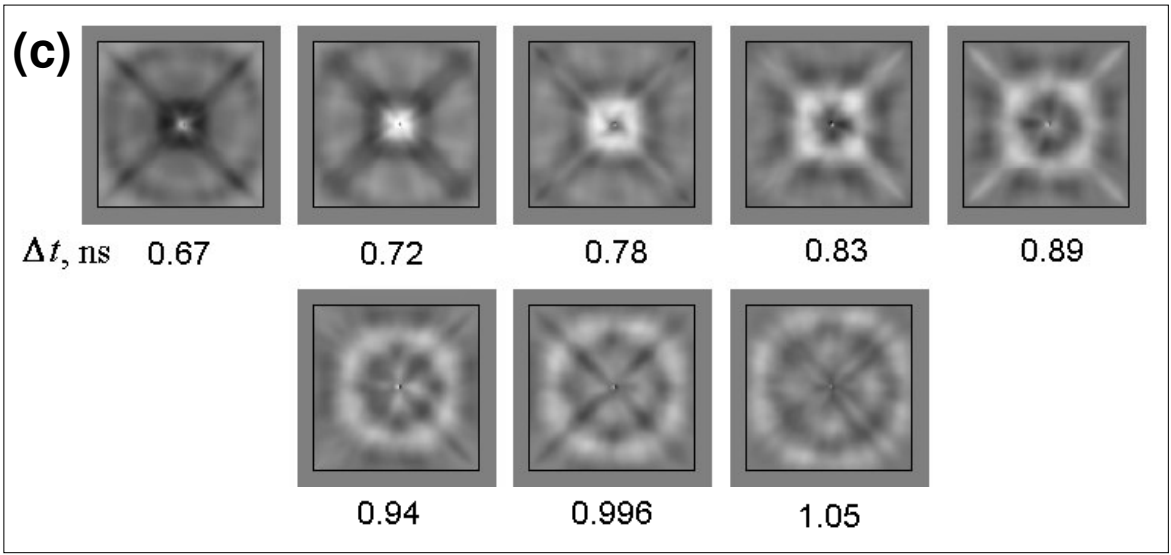

Fig. 9

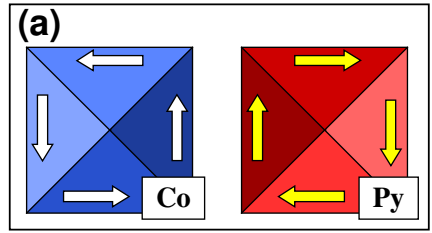

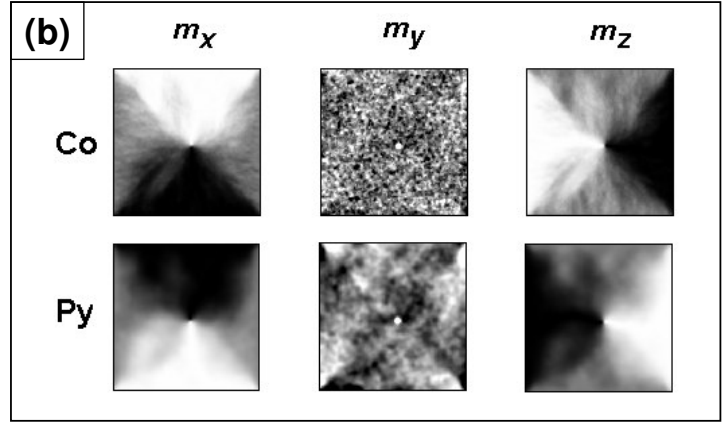

Fig. 10

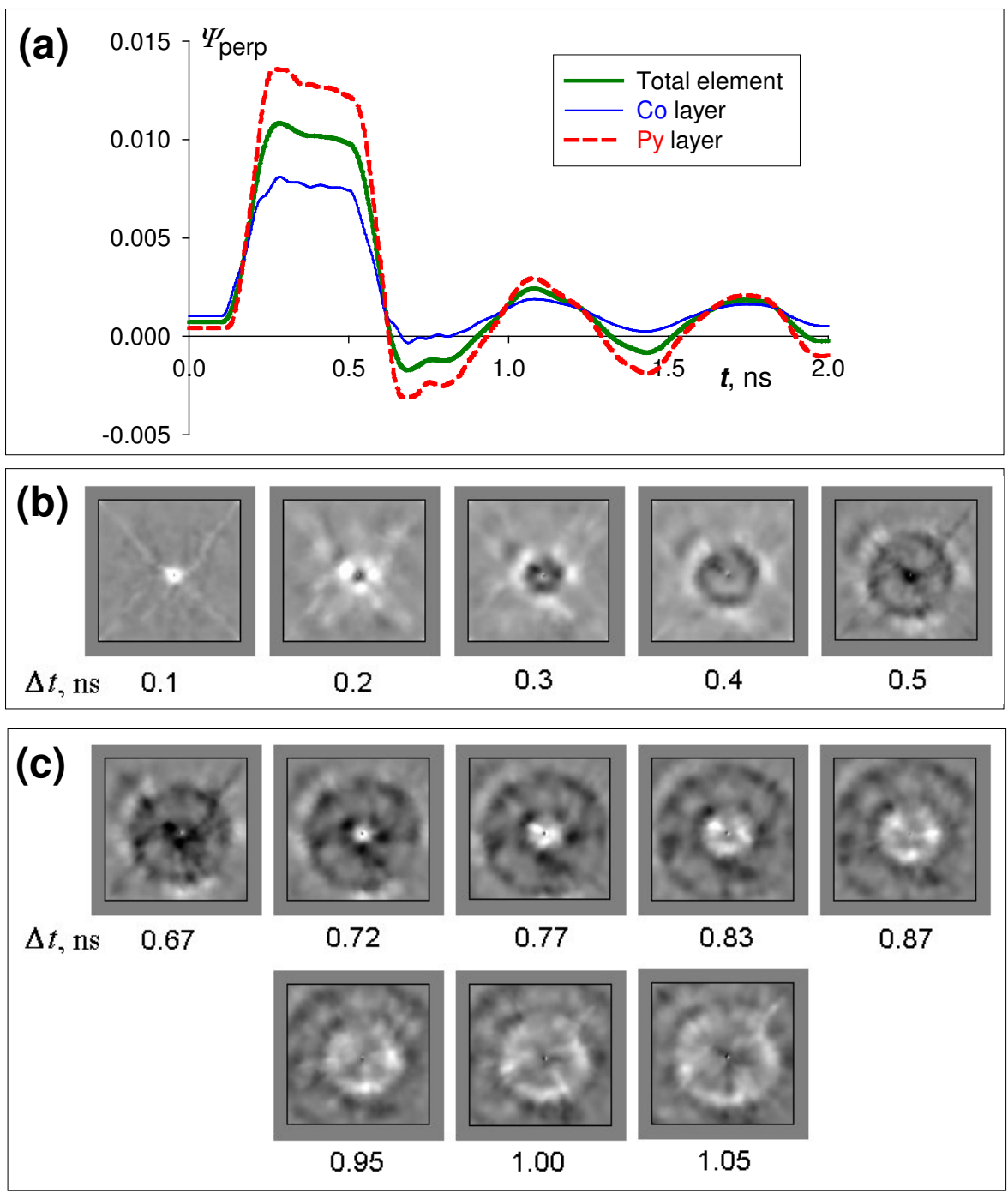

Fig. 11

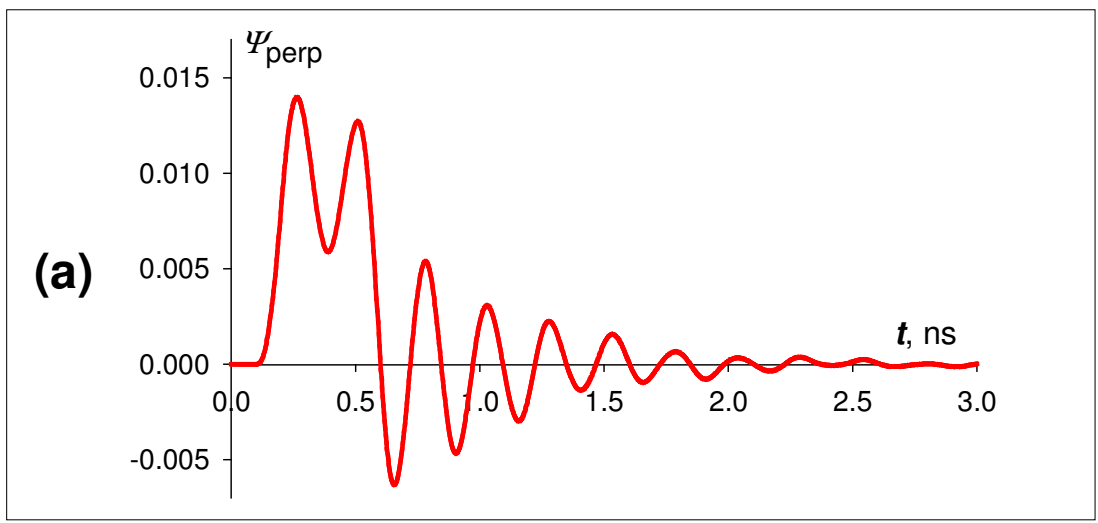
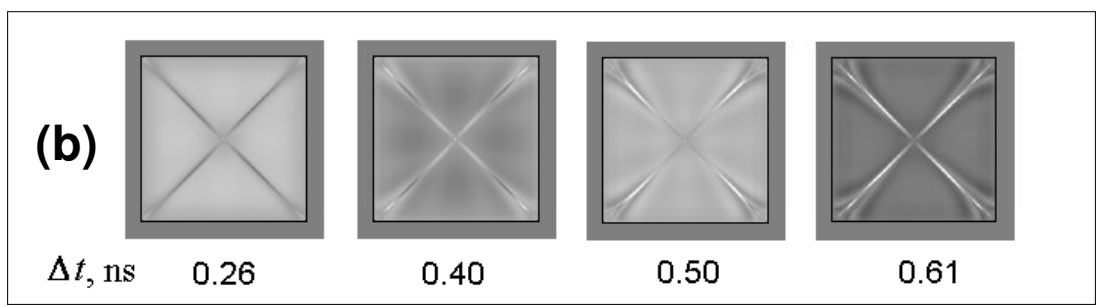
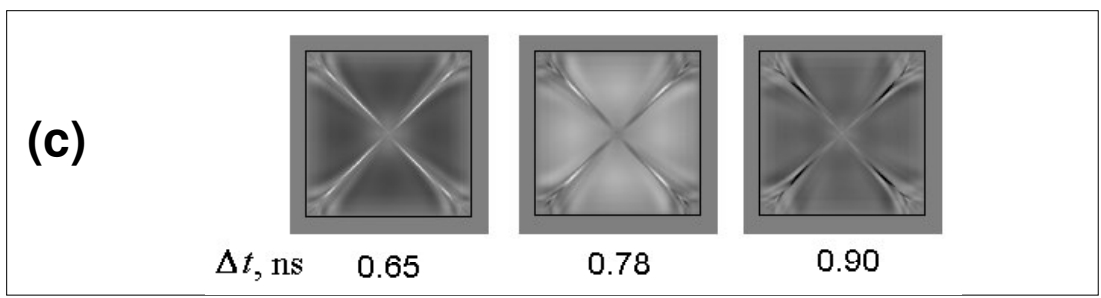

Fig. 12

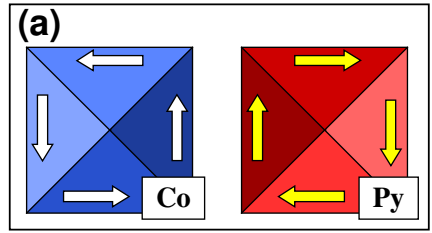

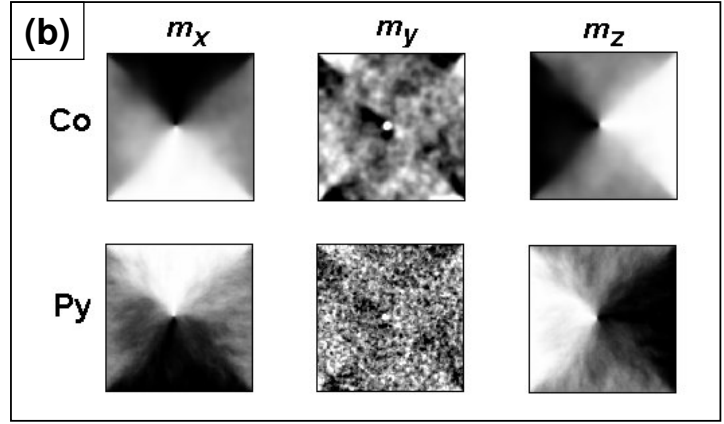

Fig. 13

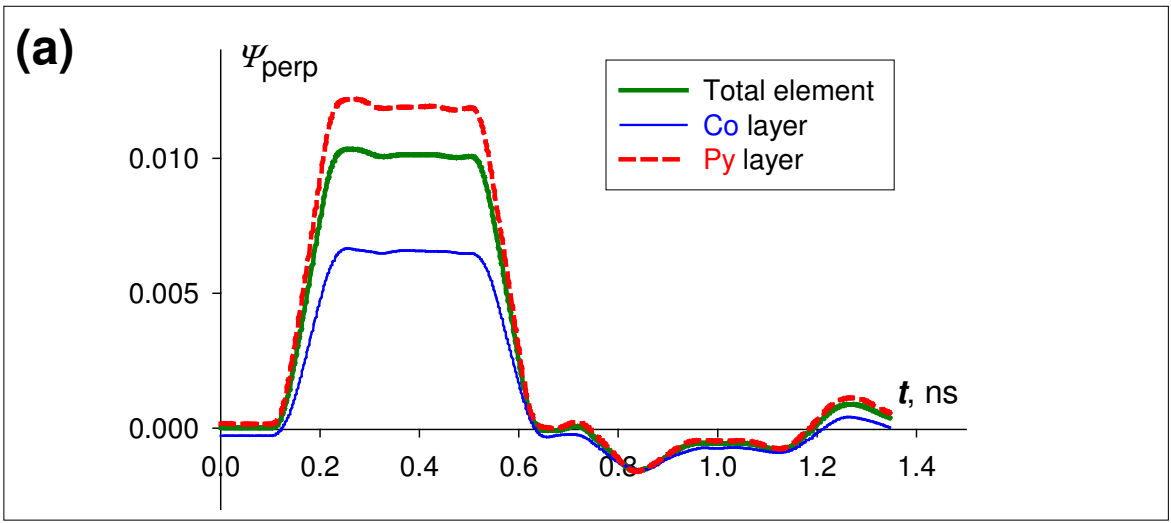
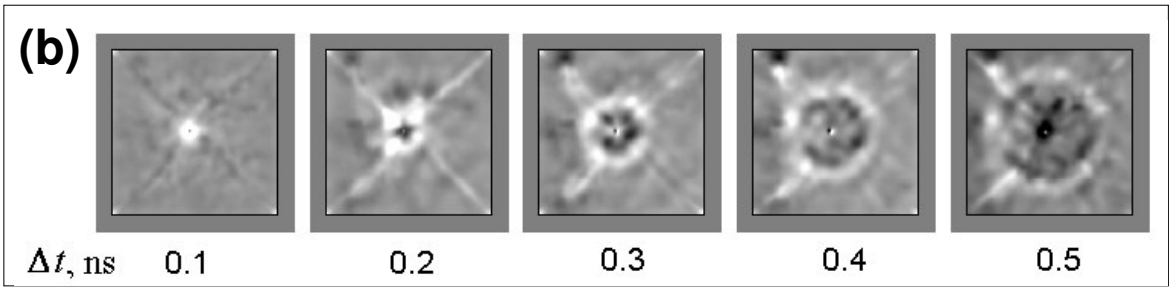
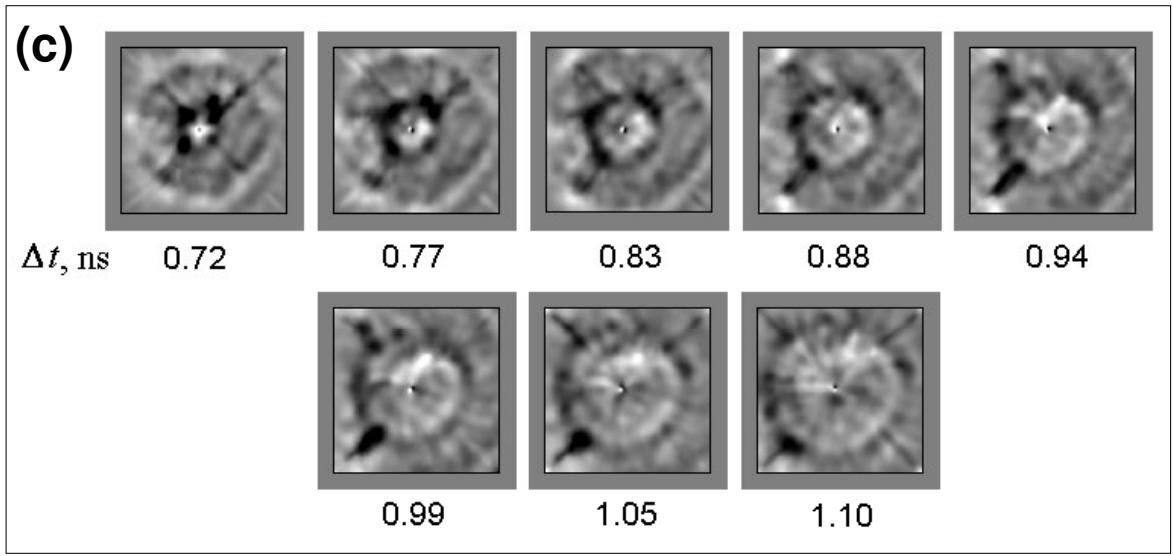

Fig. 14